\documentclass[12pt]{article}
\usepackage[toc,page]{appendix}
\usepackage{amsmath}
\usepackage{amsfonts}
\usepackage{graphicx}
\usepackage{url}
\usepackage{cite}
\title{jahmm: a tool for discretizing multiple ChIP-seq profiles}
\author{Guillaume Filion and Pol Cusc\'o}

\begin{document}
\maketitle

\section{Abstract}
Chromatin immunoprecipitation and high throughput sequencing (ChIP-seq) is the \textit{de facto} standard method to map chromatin features on genomes. The output of ChIP-seq is quantitative within a single genome-wide profile, but there is no natural way to compare experiments, which is why the data is often discretized as present/absent calls. Many tools perform this task efficiently, however they process a single input at a time, which produces discretization conflicts among replicates. Here we present the implementation of a Hidden Markov Model (HMM) using mixture negative multinomial emissions to discretize ChIP-seq profiles. The method gives meaningful discretization for a wide range of features and allows to merge datasets from different origins into a single discretized profile, which resolves discretization conflicts. A quality control step performed after the discretization accepts or rejects the discretization as a whole. The implementation of the model is called jahmm, and it is available as an R package. The source can be downloaded from \url{http://github.com/gui11aume/jahmm}.

\section{Introduction}
The discovery that genes are activated and repressed by transcription factors (proteins that regulate transcription) was the foundation of the modern theory of gene regulation \cite{pmid15950866}. More recent work on histone post-translational modifications (PTMs) showed that they play a key role in the regulation of transcription. However, the influence of transcription factors and histone PTMs on transcription is still poorly understood, in part because of the discrepancy between their behavior \textit{in vivo} and \textit{in vitro}.

Chromatin immunoprecipitation (ChIP) was the first method to address the need to analyze protein-DNA interactions in the context of the nucleus \cite{pmid2454748}. Earlier methods such as footprinting and electrophoretic mobility shift assays were invaluable in their time, but they could not guarantee that a protein of interest was present on a given sequence of the genome \textit{in vivo}. The advent of microarrays and later high throughput sequencing gave genome-wide insight into the distribution of transcription factors, but these technologies raised several statistical issues that are still not resolved today. Such methods produce a large amount of data (currently of the order of 100 million reads per run), which calls for efficient and robust analysis methods.

The constant improvement of high throughput sequencing technologies makes the comparison of experiments performed at different dates inconvenient. In addition, it is practically impossible for two laboratories to produce identical ChIP-seq results due to the high number of steps and the complexity of the protocol. For these reasons, the classical approach is to discretize ChIP-seq signals to obtain a call specifying whether the feature of interest is present or absent at every position of the genome. This process is often referred-to as ``peak finding'' in the biological literature, because transcription factors are believed to bind a single location in a large neighborhood. In practice however, ChIP-seq signals (histone PTMs in particular) often consist of wide domains extending over several Kb.

Many peak finding tools have been developed since the emergence of the ChIP-seq technology, the most popular of which are PeakFinder \cite{pmid17540862}, FindPeaks \cite{pmid18599518}, CisGenome \cite{pmid18978777}, MACS \cite{pmid18798982}, SISSRs \cite{pmid18684996}, BayesPeak \cite{pmid19772557} and HPeak \cite{pmid20598134}. BayesPeak and HPeak are based on elaborate statistical models accounting for the overdispersion of ChIP-seq signals and implement a Hidden Markov Model (HMM). However, all these tools can discretize only one ChIP-seq profile at a time, which creates call conflicts when replicates are available. The IDR (Irreproducible Discovery Rate \cite{li2011}) is an endeavour to solve this issue, but it is restricted to two replicates, meaning that there is no solution for conflict resolution when more than two replicates are available.

Here we present a model addressing this issue. The jahmm (Just Another HMM) discretizer uses an HMM with mixture negative multinomial emissions. This distribution is a good representation of the sequence count at the output of modern sequencers, and it offers an intuitive interpretation as Gamma-Poisson process. The jahmm discretizer not only allows to discretize any ChIP-seq profile, it also allows to combine signals from different sources and/or different technologies into a single discretized profile. Finally, jahmm includes an atomic quality control step that either accepts the discretization or rejects it as a whole.

\section{Results}

Here we present an accessible overview of jahmm. Mathematical details and complements can be found in the annexes.

\subsection{Motivation for the emission model}

At the output of a ChIP-seq experiment, we assume that the genome is segmented in windows of identical size and that reads from the sequencer are mapped on the genome and binned in those windows. The number of reads mapping to a genomic window is a discrete variable without upper limit, so the Poisson distribution comes as a natural first guess. However, this choice imposes that the mean number of reads is equal to the variance, which poorly matches experimental observations. It is indeed well known that the distribution of read counts in ChIP-seq experiments is overdispersed \cite{pmid19772557,pmid20598134}.

Fig. 1a shows  the read count distribution in an experiment performed without immunoprecipitation (the DNA is broken by sonication and sequenced), which describes the baseline distribution of ChIP-seq signals for 300 bp windows. The red histogram shows the distribution of a Poisson variable fitted to the observation. The variance of the observed distribution is more than 3 times larger than the mean and the difference between these distributions is evident for low read counts. For larger windows, the lack of fit of the Poisson distribution becomes more pronounced, as shown in Fig. 1b (in this case the variance is more than 10 times larger than the mean). Discarding non mappable windows reduces the skew but the resulting distribution is not Poisson (data not shown). In summary, the Poisson distribution is not suitable to model ChIP-seq experiments.

\begin{figure}
  \includegraphics[width=\textwidth]{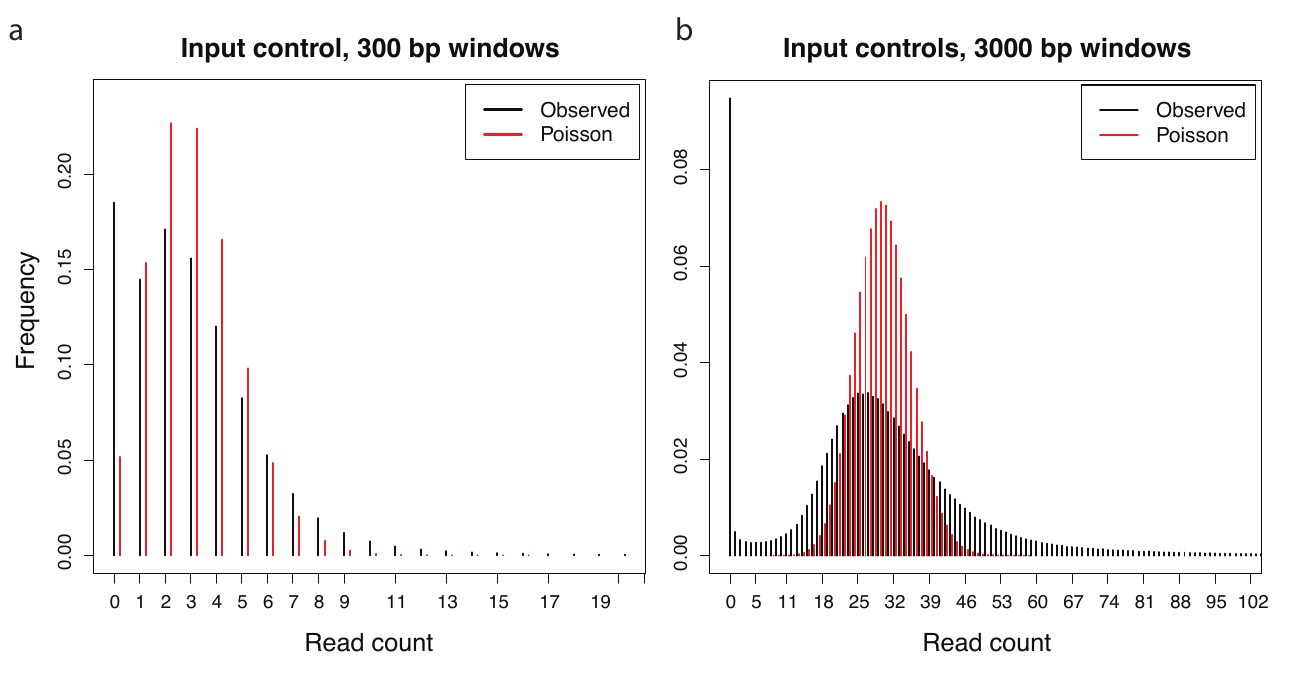}
  \caption{ChIP-seq read count distribution.
  Left (\textbf{a}): distribution of read counts for a negative control
  experiment in 300 bp windows (black bars) and the corresponding fitted
  Poisson distribution (red bars). Notice the lack of fit for the number
  of windows with no read and for windows with 7 and higher reads.
  Right (\textbf{b}): same as \textbf{a} for 3000 bp windows.}
\end{figure}

The negative binomial distribution is more flexibile because it has two parameters, which allows to separate the mean from the variance. More importantly, an intuition of this distribution is given by the two step ``Gamma-Poisson mixture''. In the first step, a parameter $\lambda$ is drawn from a Gamma distribution; in the second step, a random observation is drawn from a Poisson distribution with parameter $\lambda$. In other words, the negative binomial distribution can be viewed as a mixture of Poisson distributions with means (\textit{i.e.} $\lambda$ parameters) distributed as a Gamma random variable.

In the case of ChIP-seq experiments, the mean number of reads mapping to a window is expected to vary due to experimental and computational biases. The G+C content is known to affect the efficiency of the PCR amplification taking place before sequencing. As a consequence, the number of reads is expected to depend on the G+C content of the window. In addition, read mappability is not constant throughout the genome because of polymorphism and repeated sequences, which can decrease the number of mappable reads. These variations are not expected to have an exact Gamma distribution, but since the shape of the Gamma family is flexible, it is a good approximation for many unimodal distributions.

However, the read distribution is clearly bimodal for large windows (Fig. 1b) and is skewed for smaller windows (Fig. 1a). This bimodality is mostly due to the repeated sequences of the genome, since mapping the human genome sequence (hg19) onto itself without any experimental step yields a multimodal distribution (not shown). A mixture of two negative binomial distributions was thus chosen to model the amount of read counts mapping to each genomic window. The mixture model can be estimated efficiently with the EM algorithm \cite{Dempster77maximumlikelihood} and gives a good fit for short windows (Fig. 2a). For 3000 bp windows, the central part of the distribution shows a misfit, but the tails are well captured by the model, which makes it robust to overdispersion. Fitting the right tail is a key property for a discretization model because it reduces the number of false positives compared to the Poisson distribution.

\begin{figure}
  \includegraphics[width=\textwidth]{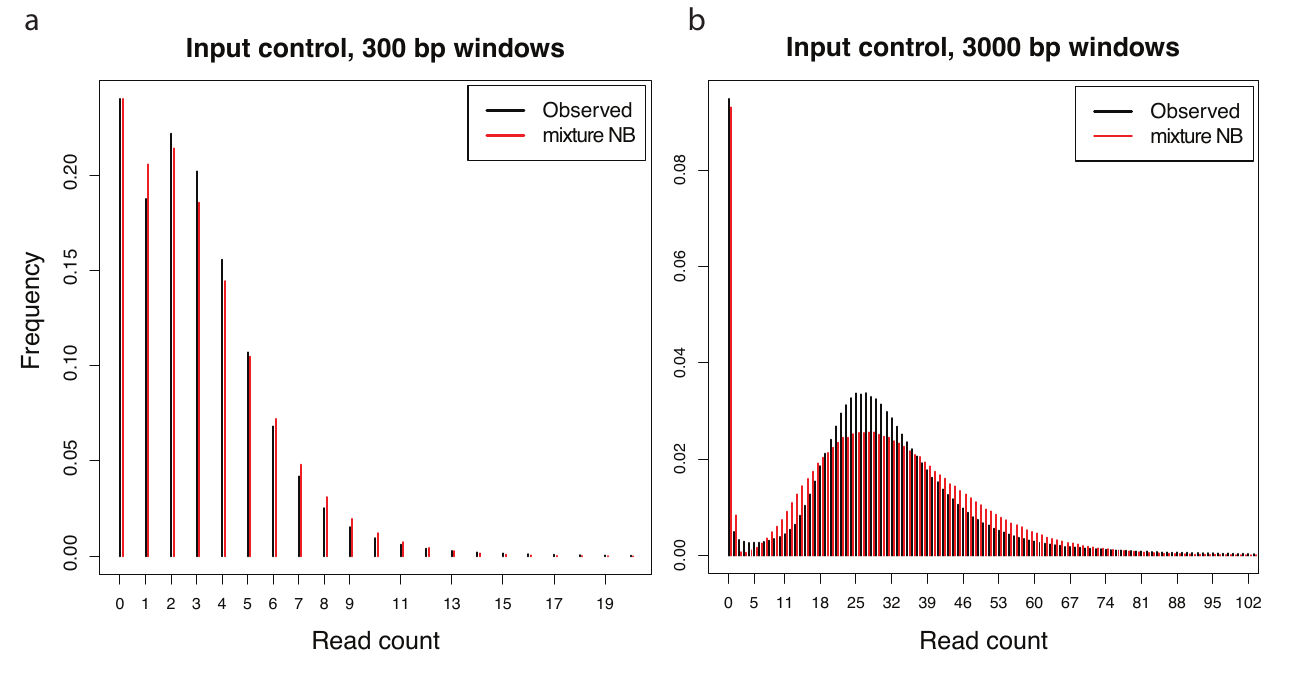}
  \caption{Fit of the mixture negative binomial model.
  Left (\textbf{a}): same as Fig. 1a, but the red bars represent the corresponding
  negative binomial mixture distribution fitted by the EM algorithm.
  Right (\textbf{b}): same as \textbf{a} for 3000 bp windows. The mixture negative
  binomial model is a good fit for the tail of the distribution.}
\end{figure}


\subsection{Implementation and test}
The input of jahmm consists of a set of binned ChIP-seq profiles (assumed to be replicates of each other) plus one negative control ChIP-seq profile binned in the same way. This profile is instrumental to estimate the baseline variations of the read count per window.
The output is a single profile of present/absent calls per genomic window. Each ChIP-seq profile represents one dimension of the emissions, modelled by the mixture negative binomial distribution motivated above. We assume that the ``shape'' parameter of the Gamma distribution underlying the Gamma-Poisson process is a global parameter fixed by the genome and the window size. This means that every genomic window is associated to a reference $\lambda$ parameter, and that the number of reads in each profile have a Poisson distribution with a fixed scaling relative to the reference. These assumptions make the profiles a mixture of negative multinomial variables.

The HMM is assumed to have 3 states, only one of which is interpreted as ``present'' or ``target''; the other 2 are interpreted as ``absent''. Hands-on experience with ChIP-seq data shows that many profiles consist of 3 distinct levels (typically ``depleted'', ``average'', ``enriched'') and that low-frequency baseline variations can sometimes capture one state of the HMM, which masks the highest peaks. For these reasons a 3-state model is more robust to process vastly different ChIP-seq data. The full model is fitted using the Baum-Welch algorithm \cite{baum1966}, followed by a multi-thread variant of simulated annealing \cite{pmid17813860} to reduce the chances of being trapped in a local optimum. The present/absent calls are then attributed to each window using the Viterbi algorithm \cite{1054010}, which returns the optimal segmentation under the observations and the fitted model.

Finally, a quality control (QC) for the segmentation is performed using the smoothing distribution of the HMM (the posterior distribution of the states given the emissions). The QC score is the estimated probability of false positives among the ``present'' calls, which expresses the confidence of the classifier for these calls. In the negative controls we have tested (profiles containing no target), the estimated false positive rate is higher than 0.09 for 300 bp windows. The QC is atomic, in other words the discretization is rejected altogether if the QC score of the sample exceeds this threshold value. Because there are high confidence peaks even in negative controls, it is more meaningful to judge the validity of the discretization, rather than the reliability of each call.

We used jahmm on ENCODE ChIP-seq data \cite{pmid22955991} for the transcription factor CTCF which is known to bind its targets as single peaks, and for the histone PTM H3K27me3 which is known to be present in the genome in domains. The datasets were produced from the K562 myelogenous leukaemia cell line by different laboratories (five distinct laboratories for CTCF and three for H3K27me3). Fig. 3a and 3b shows that the discretization closely matches the visual expectations in both cases, which is supported by the fact that the QC scores are below the rejection threshold (0.015 and 0.057 respectively).

\begin{figure}
  \includegraphics[width=\textwidth]{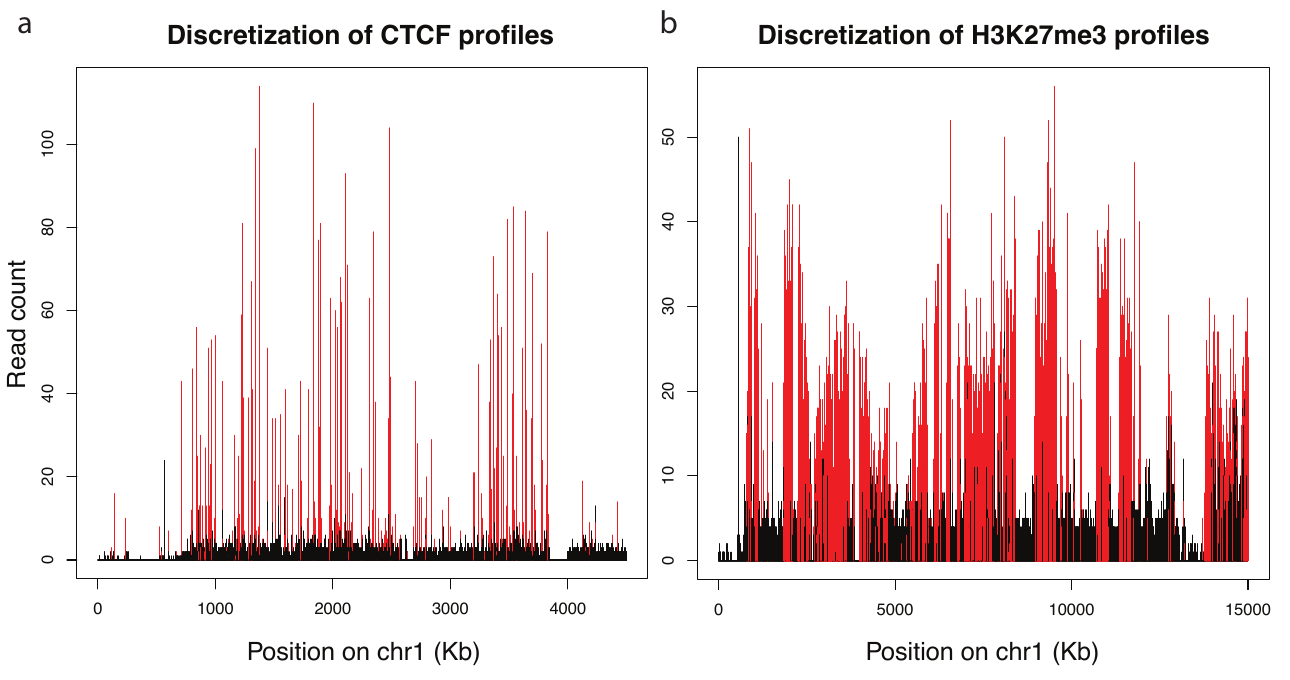}
  \caption{Example discretization by jahmm.
  Left (\textbf{a}): Discretization of CTCF binding sites. For concision
  only one of the thirteen profiles used for the discretization is shown.
  The ``present'' calls are indicated in red.
  Right (\textbf{b}): Riscretization of H3K27me3 domains.
  As for \textbf{a}, only one of the five profiles used for the discretization
  is shown with the same color code. Notice the different scale of the x axis
  in both panels.}
\end{figure}

We also used jahmm to discretize profiles of HDAC6 from a single laboratory. HDAC6 has an overwhelmingly cytosolic distribution \cite{pmid12024216}, it should therefore give a baseline signal with no target. In this case, the discretization proceeded normally, but the QC score was 0.17, exceeding the threshold. This suggests that the discretization of this profile is meaningless. Therefore jahmm can be used to discretize ChIP-seq signals of different types, without prior knowledge of the signal under study, nor of the quality of the experiment.

\section{Methods}
\subsection{ChIP-seq data processing}
The raw data .fastq files linked in the supplementary file \texttt{downloads.lst} were downloaded from the ENCODE repository.

Mapping was carried out by gem \cite{pmid23103880} with options \texttt{-q ignore -m 2 -T 4 --unique mapping}. The versions of gem-indexer and gem-mapper were 1.423 (beta), and 1.376 (beta) respectively. The
sequence of the human genome (hg19) in fasta format was downloaded from \url{http://hgdownload.cse.ucsc.edu/goldenPath/hg19/bigZips/chromFaMasked.tar.gz}.

\bibliography{references}{}

\begin{thebibliography}{10}

\bibitem{baum1966}
Leonard~E. Baum and Ted Petrie.
\newblock Statistical inference for probabilistic functions of finite state
  markov chains.
\newblock {\em The Annals of Mathematical Statistics}, 37(6):1554--1563, 12
  1966.

\bibitem{Dempster77maximumlikelihood}
A.~P. Dempster, N.~M. Laird, and D.~B. Rubin.
\newblock Maximum likelihood from incomplete data via the em algorithm.
\newblock {\em JOURNAL OF THE ROYAL STATISTICAL SOCIETY, SERIES B},
  39(1):1--38, 1977.

\bibitem{pmid18599518}
Anthony~P Fejes, Gordon Robertson, Mikhail Bilenky, Richard Varhol, Matthew
  Bainbridge, and Steven J~M Jones.
\newblock {FindPeaks 3.1: a tool for identifying areas of enrichment from
  massively parallel short-read sequencing technology.}
\newblock {\em {Bioinformatics}}, 24(15):1729--30, August 2008.

\bibitem{pmid12024216}
Charlotte Hubbert, Amaris Guardiola, Rong Shao, Yoshiharu Kawaguchi, Akihiro
  Ito, Andrew Nixon, Minoru Yoshida, Xiao-Fan Wang, and Tso-Pang Yao.
\newblock {HDAC6 is a microtubule-associated deacetylase.}
\newblock {\em {Nature}}, 417(6887):455--8, May 2002.

\bibitem{pmid18978777}
Hongkai Ji, Hui Jiang, Wenxiu Ma, David~S Johnson, Richard~M Myers, and Wing~H
  Wong.
\newblock {An integrated software system for analyzing ChIP-chip and ChIP-seq
  data.}
\newblock {\em {Nat. Biotechnol.}}, 26(11):1293--300, November 2008.

\bibitem{pmid17540862}
David~S Johnson, Ali Mortazavi, Richard~M Myers, and Barbara Wold.
\newblock {Genome-wide mapping of in vivo protein-DNA interactions.}
\newblock {\em {Science}}, 316(5830):1497--502, June 2007.

\bibitem{pmid18684996}
Raja Jothi, Suresh Cuddapah, Artem Barski, Kairong Cui, and Keji Zhao.
\newblock {Genome-wide identification of in vivo protein-DNA binding sites from
  ChIP-Seq data.}
\newblock {\em {Nucleic Acids Res.}}, 36(16):5221--31, September 2008.

\bibitem{pmid17813860}
S~Kirkpatrick, C~D Gelatt, and M~P Vecchi.
\newblock {Optimization by simulated annealing.}
\newblock {\em {Science}}, 220(4598):671--80, May 1983.

\bibitem{pmid22955991}
Stephen~G Landt, Georgi~K Marinov, Anshul Kundaje, Pouya Kheradpour, Florencia
  Pauli, Serafim Batzoglou, Bradley~E Bernstein, Peter Bickel, James~B Brown,
  Philip Cayting, Yiwen Chen, Gilberto DeSalvo, Charles Epstein, Katherine~I
  Fisher-Aylor, Ghia Euskirchen, Mark Gerstein, Jason Gertz, Alexander~J
  Hartemink, Michael~M Hoffman, Vishwanath~R Iyer, Youngsook~L Jung, Subhradip
  Karmakar, Manolis Kellis, Peter~V Kharchenko, Qunhua Li, Tao Liu, X~Shirley
  Liu, Lijia Ma, Aleksandar Milosavljevic, Richard~M Myers, Peter~J Park,
  Michael~J Pazin, Marc~D Perry, Debasish Raha, Timothy~E Reddy, Joel Rozowsky,
  Noam Shoresh, Arend Sidow, Matthew Slattery, John~A Stamatoyannopoulos,
  Michael~Y Tolstorukov, Kevin~P White, Simon Xi, Peggy~J Farnham, Jason~D
  Lieb, Barbara~J Wold, and Michael Snyder.
\newblock {ChIP-seq guidelines and practices of the ENCODE and modENCODE
  consortia.}
\newblock {\em {Genome Res.}}, 22(9):1813--31, September 2012.

\bibitem{li2011}
Qunhua Li, James~B. Brown, Haiyan Huang, and Peter~J. Bickel.
\newblock Measuring reproducibility of high-throughput experiments.
\newblock {\em The Annals of Applied Statistics}, 5(3):1752--1779, 09 2011.

\bibitem{pmid23103880}
Santiago Marco-Sola, Michael Sammeth, Roderic Guigó, and Paolo Ribeca.
\newblock {The GEM mapper: fast, accurate and versatile alignment by
  filtration.}
\newblock {\em {Nat. Methods}}, 9(12):1185--8, December 2012.

\bibitem{pmid15950866}
Mark Ptashne.
\newblock {Regulation of transcription: from lambda to eukaryotes.}
\newblock {\em {Trends Biochem. Sci.}}, 30(6):275--9, June 2005.

\bibitem{pmid20598134}
Zhaohui~S Qin, Jianjun Yu, Jincheng Shen, Christopher~A Maher, Ming Hu, Shanker
  Kalyana-Sundaram, Jindan Yu, and Arul~M Chinnaiyan.
\newblock {HPeak: an HMM-based algorithm for defining read-enriched regions in
  ChIP-Seq data.}
\newblock {\em {BMC Bioinformatics}}, 11:369, 2010.

\bibitem{pmid2454748}
M~J Solomon, P~L Larsen, and A~Varshavsky.
\newblock {Mapping protein-DNA interactions in vivo with formaldehyde: evidence
  that histone H4 is retained on a highly transcribed gene.}
\newblock {\em {Cell}}, 53(6):937--47, June 1988.

\bibitem{pmid19772557}
Christiana Spyrou, Rory Stark, Andy~G Lynch, and Simon Tavaré.
\newblock {BayesPeak: Bayesian analysis of ChIP-seq data.}
\newblock {\em {BMC Bioinformatics}}, 10:299, 2009.

\bibitem{1054010}
A.J. Viterbi.
\newblock Error bounds for convolutional codes and an asymptotically optimum
  decoding algorithm.
\newblock {\em Information Theory, IEEE Transactions on}, 13(2):260--269, April
  1967.

\bibitem{pmid18798982}
Yong Zhang, Tao Liu, Clifford~A Meyer, Jérôme Eeckhoute, David~S Johnson,
  Bradley~E Bernstein, Chad Nusbaum, Richard~M Myers, Myles Brown, Wei Li, and
  X~Shirley Liu.
\newblock {Model-based analysis of ChIP-Seq (MACS).}
\newblock {\em {Genome Biol.}}, 9(9):R137, 2008.

\end{thebibliography}
\bibliographystyle{plain}


\newpage
\begin{appendices}

In the text, we often refer to the digamma and trigamma
functions. The digamma function, noted $\psi(\alpha)$ is the
derivative of $\log \Gamma(\alpha)$, and the trigramma function,
noted $\psi'(\alpha)$ is the derivative of the digamma function.

\section{The negative multinomial distribution}
\label{nb}

\subsection{The Gamma-Poisson approach}
\label{subsection_gamma_poisson}

    In what follows, $y$ is a non negative integer (an element of
    $\mathbb{N}$). Let $Y$ be a
    discrete random variable distributed according to the Poisson
    distribution with parameter $\lambda$, denoted $P(\lambda)$.
    The probability that $Y$ is equal to $y$ is by definition

    $$ P(Y=y) = e^{-\lambda} \frac{\lambda^y}{y!}. $$

    Let us now assume that $\lambda$ is itself a random variable,
    such that the above equality is actually $P(Y=y | \lambda)$.
    If $\lambda$ has a Gamma distribution with parameters $\alpha$
    and $\beta$, the joint distribution of $Y$ and $\lambda$ is
    written as

    $$ P(Y=y, \lambda) = e^{-\lambda} \frac{\lambda^y}{y!}
         \frac{1}{\Gamma(\alpha)\beta^{\alpha}} e^{-\lambda/\beta}
         \lambda^{\alpha-1}. $$

    The marginal distribution of $Y$, \textit{i.e.} $P(Y=y)$, is found
    by integrating the equality above over $\lambda$.

    \begin{align}
      P(Y=y) &= \frac{1}{\Gamma(\alpha)\beta^{\alpha}y!}
         \int_0^{+\infty} e^{-\lambda(1+1/\beta)} \lambda^{\alpha+y-1}
         d\lambda \nonumber \\
        &= \frac{\Gamma(\alpha+y)}{\Gamma(\alpha)\beta^{\alpha}
          (1+1/\beta)^{\alpha+y} y!} \nonumber \\
        &= \frac{\Gamma(\alpha+y)}{\Gamma(\alpha)y!}
          \left(\frac{1}{1+\beta}\right)^{\alpha}
          \left(\frac{\beta}{1+\beta}\right)^y.
\label{NBdistrib}
    \end{align}

    Equation (\ref{NBdistrib}) is the expression of the negative
    binomial distribution, with a somewhat unusual parametrization.
    We will refer to this distribution as a negative binomial
    with parameters $(\alpha, 1/(1+\beta))$.

  \subsection{The negative multinomial distribution}

    As introduced in section \ref{subsection_gamma_poisson},
    the following equation defines $r$ Poisson
    variables that are conditionally independent given $\lambda$

    \begin{equation}
\label{bayesian_nm}
      P(Y_1=y_1, \ldots, Y_r=y_r|\lambda) = 
      e^{-\gamma_1\lambda}\frac{(\gamma_1\lambda)^{y_1}}{y_1!}
      \times \ldots \times
      e^{-\gamma_r\lambda}\frac{(\gamma_r\lambda)^{y_r}}{y_r!}.
    \end{equation}

    Multiplying by the density of $\lambda$ and integrating as
    above, the marginal distribution of the vector
    $(Y_1, \ldots, Y_r)$ comes out to

    \begin{align}
\label{NMdistrib}
      P(Y_1=y_1, \ldots, Y_r=y_r) &=
      \frac{\Gamma(\alpha+y_1+\ldots+y_r)}
      {\Gamma(\alpha)y_1!\ldots y_r!}p_0^{\alpha}p_1^{y_1}
      \ldots p_r^{y_r}, \; \text{where}                \\
      p_0 &= \frac{1/\beta}{1/\beta+\gamma_1+\ldots+\gamma_r},
      \; \text{and} \nonumber \\
      p_i &= \frac{\gamma_i}{1/\beta+\gamma_1+\ldots+\gamma_r},
      \; \text{for} \; i = 1, \ldots, r. \nonumber
    \end{align}

    This distribution is called the negative multinomial. We will
    refer to it is as a negative multinomial with parameters
    $(\alpha, p_1, \ldots, p_r)$.
    We have shown that it can be interpreted as the observations of
    a Gamma-Poisson process, where a common $\lambda$ value is
    drawn from a Gamma distribution, and $r$ variables are drawn
    from independent Poisson distributions with scalings
    $\gamma_1, \ldots, \gamma_r$ relative to $\lambda$. Note that
    the variables $Y_1, \ldots, Y_r$ are independent contionally on
    $\lambda$, but in section \ref{subsection_marginal_nm} we prove
    that they are never unconditionally independent.

    The parameters of the negative binomial distribution have
    an alternative interpretation which emphasizes their dependence.
    Suppose an urn contains black balls and balls of $r$ different
    colors in respective proportions $p_0, p_1, \ldots, p_r$. Let
    us draw balls with replacement from this urn until we draw
    a black ball for the $k$-th time, and count how many balls of
    each color we drew. The probability of the $r$-tuple
    $(y_1, \ldots, y_r)$ is easily seen to be

    \begin{equation*}
    {k-1+y_1+\ldots+y_r \choose (k-1), y_1, \ldots, y_r}
      p_0^k p_1^{y_1} \ldots p_r^{y_r} =
    \frac{\Gamma(k+y_1+\ldots+y_r)}{\Gamma(k)y_1! \ldots y_r!}
      p_0^k p_1^{y_1} \ldots p_r^{y_r}.
    \end{equation*}

    This is formula (\ref{NMdistrib}), where $\alpha$ has
    been replaced by $k$. The negative multinomial distribution is
    a generalization of the drawing process described above with
    non integer values of $k$. The ball and urn interpretation
    makes it clear that the observed counts $(y_1, \ldots, y_r)$
    are expected to be twice smaller for a twice larger value of
    $p_0$ or for a twice smaller value of $\alpha$.

\subsection{Marginal distributions}
\label{subsection_marginal_nm}

    Finally, we compute the marginal distributions of
    $(Y_1, \ldots, Y_r)$. Summing (\ref{NMdistrib}) is straightforward,
    but instead we observe that taking the margins of (\ref{bayesian_nm})
    and integrating over $\lambda$ as above yields for
    $l = 1, \ldots, r$

    \begin{align*}
    P(Y_l = y_l) &= \frac{\Gamma(\alpha+y_l)}{\Gamma(\alpha)y_l!}
      p_0^{*\alpha} p_l^{*y_l}, \; \text{where} \\
    p_0^* &= \frac{1/\beta}{1/\beta + \gamma_l}, \; \text{and} \\
    p_l^* &= \frac{\gamma_l}{1/\beta + \gamma_l}.
    \end{align*}
    
    Not surprisingly, we obtain a negative binomial distribution.
    More interestingly though, the parameters of this
    distribution are linked to the previous parameters
    by the equality $p_0^* / p_l^* = p_0 / p_l$. These constraints
    are valid for any number of variables in the negative
    multinomial model, so they come in handy to reparametrize the
    model every time variables are added or dropped.

    As an example of the use of these constraints, we 
    show with $r=2$ that the margins of a negative multinomial
    distribution are never independent (for the general case,
    observe that mutual independence entails pairwise independence
    and that the margins over $r-2$ variables have a negative
    multinomial distribution). Let us fix $z_2 = 0$. The terms
    $P(z_1 = k, z_2 = 0)$ are proportional to
    $\Gamma(\alpha+k) p_1^k/k!$ and the terms $P(z_1 = k)P(z_2 = 0)$
    are proportional to $\Gamma(\alpha+k) p_1^{*k}/k!$ where
    $p_1^* = p_1 / (p_1 + p_0) < p_1$ so equality cannot hold
    for every $k \geq 0$. This shows that the joint distribution
    is never equal to the product of the marginal distributions.

    Note that the proof above assumes $p_0 > 0$, which is
    a consequence of $\beta < \infty$. So as long as $\lambda$ is 
    distributed according to a proper Gamma distribution, which
    is a defining feature of the negative multinomial distribution, the
    variables cannot be independent.

    From the marginal distributions we can compute the conditional
    distribution of $(Y_1, \ldots, Y_i)$ given $(Y_{i+1}, \ldots, Y_r)$
    (and similary the distribution of any set of variables given
    the complentary set). Using the same rationale as above, the
    marginal distribution is found to be negative multinomial with

    \begin{align*}
    P(Y_{i+1}=y_{i+1}, \ldots, Y_r=y_r) &= \\
      \frac{\Gamma(\alpha + y_{r+1} + \ldots + y_r)}
      {\Gamma(\alpha)y_{i+1}! \ldots y_r!} &p_0^{\alpha}p_{i+1}^{y_{i+1}}
      \ldots p_r^{y_r} \left(\frac{1}{p_0 + p_{i+1} + \ldots + p_r}
      \right)^{\alpha+y_{i+1} + \ldots + y_r}.
    \end{align*}

    The conditional distribution is computed as the ratio of the
    full distribution and the marginal distribution.

    \begin{align*}
      P(Y_1=y_1, \ldots, Y_i=y_i|Y_{i+1}=y_{i+1}, \ldots, Y_r=y_r) &= \\
      \frac{\Gamma(\alpha + y_1 + \ldots + y_r)}
      {\Gamma(\alpha+y_{i+1}+\ldots+y_r)y_1! \ldots y_i!}
      &q_0^{\alpha+y_{i+1}+\ldots+y_r}p_1^{y_1} \ldots p_i^{y_i},
    \end{align*}

    \noindent
    where $q_0 = 1-(p_1+\ldots+p_i)$. In other words, the
    distribution of $(Y_1, \ldots, Y_i)$ given $(Y_{i+1}, \ldots, Y_r)$
    is negative multinomial with parameters
    $(\alpha+y_{i+1}+\ldots+y_r, p_1, \ldots, p_i)$.

    \section{Hidden Markov models}

    We will consider only discrete Hidden Markov models (HMMs) and
    will simply refer to them as Hidden Markov model, without mention
    of the term `discrete' for simplicity.
    HMMs are defined by 

    \begin{enumerate}
      \item a set $S$ of $m$ states numbered from 1 to $m$,
      \item an initial state probability distribution $\nu$, which
      gives the probabilities that the system is initially in state $i$,
      \item an $m \times m$ transition matrix $Q$ which contains the
      probabilities $Q(i,j)$ that the system goes from state $i$ to
      state $j$,
      \item $m$ distributions denoted $g_i$ $(i = 1, \ldots, m)$, which
      give the emission probabilities in the different states.
    \end{enumerate}

    \subsection{The Forward-Backward algorithm}

    For a sequence of emissions $y_0, \ldots, y_n$, the likelihood
    of the state sequence $i_0, \ldots, i_n$ is proportional to

    $$ \nu(i_0)g_{i_0}(y_0)
       \prod_{k=1}^n Q(i_{k-1},i_k)g_{i_k}(y_k). $$

    By summing over all possible combinations of states, we obtain
    the normalizing constant $L_n$ such that

    \begin{equation}
       L_n = \sum_{i_0 \in S, \ldots, i_n \in S} \nu(i_0)g_{i_0}(y_0)
       \prod_{k=1}^n Q(i_{k-1},i_k)g_{i_k}(y_k).
    \end{equation}

    We denote $\phi_{k|n}(i)$ the probability that the system is in
    state $i$ at time $k$ given the emissions $y_0, \ldots, y_n$. If
    we call $S_n(k,i)$ the set of $n$-tuples $(i_0, \ldots, i_n)$
    such that $i_k = i$, the value of $\phi_{k|n}(i)$ comes as

    $$ \phi_{k|n}(i) = \frac{1}{L_n}
       \sum_{(i_0, \ldots, i_n) \in S_n(k,i)}
       \nu(i_0)g_{i_0}(y_0) \prod_{l=1}^n Q(i_{l-1}, i_l)
       g_{i_l}(y_l). $$

    We now introduce $\alpha_k(i)$ the probability that the
    system is in state $i$ at time $k$ given the emissions
    $y_0, \ldots, y_k$, and the $\beta_{k|n}(\cdot)$ the numerical
    function such that $\phi_{k|n}(i) = \alpha_k(i)\beta_{k|n}(i)$.
    
    \begin{align*}
      \alpha_k(i) &= \frac{1}{L_k}
      \sum_{i_0=1}^m \cdots \sum_{i_{k-1}=1}^m
      \nu(i_0)g_{i_0}(y_0) \prod_{l=1}^{k-1} Q(i_{l-1},i_l) g_{i_l}(y_l)
      Q(i_{k-1}, i)g_i(y_k) \\
      \beta_{k|n}(i) &= \frac{L_k}{L_n}
      \sum_{i_{k+1}=1}^m \cdots \sum_{i_n=1}^m
      Q(i, i_{k+1})g_{i_{k+1}}(y_{k+1})
      \prod_{l=k+2}^n Q(i_{l-1}, i_l)g_{i_l}(y_l)
    \end{align*}

    To preserve the equality $\phi_{k|n}(i) = \alpha_k(i)\beta_{k|n}(i)$
    for every $k$, we set by definition $\beta_{n|n}(i) = 1$.
    From the equations above, we draw the following recursive
    equations:

    \begin{align} \alpha_k(i) &= \frac{L_{k-1}}{L_k}
      \sum_{j=1}^m \alpha_{k-1}(j) Q(j,i) g_i(y_k) \label{alpha} \\
      \beta_{k|n}(i) &= \frac{L_k}{L_{k+1}}
      \sum_{j=1}^m Q(i,j) g_j(y_{k+1})
      \beta_{k+1|n}(j). \label{beta}
    \end{align}

    Equations (\ref{alpha}) and (\ref{beta}) are the basis of the
    Forward-Backward algorithm to compute $\phi_{k|n}(i)$. The terms
    $\alpha_k(i)$ can be recursively computed from $k=0$ to $k=n$
    with equation (\ref{alpha}), and the terms $\beta_{k|n}(i)$
    can be computed from $k=n-1$ to $k=0$ with equation (\ref{beta}).
    The terms $\phi_{k|n}(i)$ are then found as the product
    $\alpha_k(i)\beta_{k|n}(i)$.

    We now turn to the term $\phi_{k-1,k|n}(i,j)$, which is by
    definition the probability that the system is in state $i$ at
    time $k-1$ and in state $j$ at time $k$ given $y_0, \ldots, y_n$.
    If we call $S_n(k,i,j)$
    the set of $n$-tuples $(i_0, \ldots, i_n)$ such that
    $i_{k-1} = i$ and $i_k = j$, we get

    \begin{align}
      \phi_{k-1,k|n}(i,j) &= \frac{1}{L_n}
       \sum_{(i_0, \ldots, i_n) \in S_n(k,i,j)}
       \nu(i_0)g_{i_0}(y_0) \prod_{l=1}^n Q(i_{l-1}, i_l)
       g_{i_l}(y_l) \nonumber \\
        &= \frac{L_{k-1}}{L_k}
       \alpha_{k-1}(i) Q(i,j) g_j(y_k) \beta_k(j). \label{phiQ}
    \end{align}

    When the $\alpha_k(i)$ and the $\beta_{k|n}(i)$ have been
    computed by the Forward-Backward algorithm, we also have access
    to the $\phi_{k-1,k|n}(i,j)$ by using formula (\ref{phiQ}).

\subsection{The Baum-Welch algorithm}

    The Baum-Welch algorithm is the special case of the EM algorithm
    applied to HMMs. Let us consider the general case of the triplet
    $(X, Z, \theta)$ where the variable $X$ is observed, $Z$ is not
    observed, and $\theta$ is the set of parameters of the distribution
    of $(X,Z)$. The full likelihood $\mathcal{L}_0(X, Z, \theta)$
    cannot be computed because the value of $Z$ is unknown.

    To find the value of $\theta$ that maximizes the full likelihood,
    we introduce an iterative procedure where the values of the
    parameter are updated upon each iteration. The current value of
    $\theta$ is noted $\theta^{(t)}$, and we compute the expected
    complete log-likelihood $\mathcal{Q}(\theta|\theta^{(t)})$
    assuming the current value of $\theta$ (note the difference
    between the intermediate quantity of the EM $\mathcal{Q}$ and
    the transition matrix $Q$).

    $$ \mathcal{Q}(\theta|\theta^{(t)}) =
      E_{Z|X, \theta^{(t)}} \left\{
      \log \mathcal{L}_0(X, Z, \theta^{(t)}) \right\}$$

    This computation is called the E-step. The notations mean that
    the expectation is taken over the variable $Z$, assuming that
    it is conditional on the observed values of $X$ and that the
    parameters of the distribution are given by $\theta^{(t)}$.
    The E-step is followed by the
    M-step, in which $\theta^{(t+1)}$ is set to the value of
    $\theta$ that maximizes $\mathcal{Q}(\theta|\theta^{(t)})$.

    In the case of HMMs, the variable that is not observed is the
    sequence of states. The set of parameters $\theta^{(t)}$
    represents the transition probabilities (the matrix $Q$)
    and the parameters of the $m$ distributions of the emissions.

    The log-likelihood of the state sequence $(i_0, \ldots, i_n)$
    is

    $$ \log \nu(i_0) + \sum_{k=1}^n \log Q(i_{l-1}, i_l)
      + \sum_{k=0}^n \log g_{i_l}(i_l, \theta). $$

    The addition of $\theta$ to the terms above emphasizes that they
    depend on the value of the parameters. To compute
    $\mathcal{Q}(\theta|\theta^{(t)})$, we need to take the
    expectation of the above over the state sequence
    conditionally on $y_0, \ldots, y_n$ and assuming that the parameters
    are given by $\theta^{(t)}$.

    \begin{align}
      \mathcal{Q}(\theta|\theta^{(t)}) &=
      E_{\theta^{(t)}} \left\{ \log \nu(i_0)
      \big| y_0, \ldots, y_n \right\} + \nonumber \\
      &\sum_{k=1}^n E_{\theta^{(t)}} \left\{
        \log Q(i_{l-1}, i_l)\big| y_0,
            \ldots, y_n \right\} + \label{QEM} \\
      &\sum_{k=0}^n E_{\theta^{(t)}} \left\{
        \log g_{i_l}(y_l, \theta) \big| y_0, \ldots, y_n \right\}
        \nonumber
    \end{align}

    In practice, the first term of (\ref{QEM}) will often not depend
    on $\theta$ so it will not contribute to the evaluation.
    The third term can be rewritten as

    $$\sum_{k=0}^n\sum_{i=1}^m \phi_k(i) \log g_{i_l}(y_l, \theta).$$

    This term depends on the emission probabilities, and nothing
    can be said about it in general terms because they differ
    between different models. But the second term depends only on
    the transition probabilities, which are present in every HMM,
    and it can be solved in general. First we notice that

    \begin{align*}
      &E_{\theta^{(t)}} \left\{ \log Q(i_{l-1}, i_l)
      \big| y_0, \ldots, y_n \right\} = \\
      &E_{\theta^{(t)}} \left\{ \sum_{i=1}^m\sum_{j=1}^m 
      1_{\{(i_{l-1}, i_l) = (i,j)\}} \log Q(i,j)
      \big| y_0, \ldots, y_n \right\} = \\
      &\sum_{i=1}^m\sum_{j=1}^m E_{\theta^{(t)}} \left\{
      1_{\{(i_{l-1}, i_l) = (i,j)\}} \big| y_0, \ldots, y_n \right\}
      \log Q(i,j)
    \end{align*}

    Remember that by definition $E_{\theta^{(t)}} \left\{
    1_{\{(i_{l-1}, i_l) = (i,j)\}} \big| y_0, \ldots, y_n \right\}$
    is $\phi_{k-1,k}(i,j)$, so that we can rewrite the second term
    of (\ref{QEM}) as

    $$ \sum_{k=1}^n\sum_{i=1}^m\sum_{j=1}^m \phi_{k-1,k}(i,j)
      \log Q(i,j). $$

    The values of $\phi_{k-1,k}(i,j)$ are computed during the
    E-step by the Forward-Backward algorithm.
    The terms $Q(i,j)$ are part of $\theta$ and are thus updated
    during the M-step. By using Lagrange multipliers, we can show
    that the update values are

    $$ Q(i,j)^{(t+1)} = \frac{\sum_{k=1}^n \phi_{k-1,k}(i,j)}
      {\sum_{k=1}^n\sum_{l=1}^m\phi_{k-1,k}(i,l)}. $$
    
    To complete the Baum-Welch algorithm, we need to compute the last
    term of (\ref{QEM}), which requires making a model for the
    emissions.
    
    \section{Negative multinomial emissions}

    The readout of ChIP-seq and similar experiments is a sequence of
    reads mapped to genomic windows of identical size.
    The negative multinomial distribution is a good choice\footnote{One
    of the main weaknesses of that model is that it
    assumes that the distribution of the parameter $\lambda$ is IID
    for all genomic windows. This is probably not the case, as for
    every profile we expect that two neighboring windows have similar
    expected read counts.} to describe
    the number of reads per window for the following reasons:

    \begin{enumerate}
      \item it is a discrete random variables with values in
        $\mathbb{N}$.
      \item section \ref{nb} shows that it can be interpreted as a
        Poisson distribution where the parameter $\lambda$ varies
        as a Gamma variable. With this interpretation, each
        genomic window has a different expected read number.
        Conditionally on that number, the read count for a given
        window is a Poisson variable.
    \end{enumerate}

    We further assume that $r$ experiments are available.
    For a given genomic window and a given state $x_i$, the
    probability of observing $(z_1, \ldots, z_r)$ reads in the
    available profiles is

    \begin{align*}
      \frac{\Gamma(\alpha+z_1+\ldots+z_r)}
      {\Gamma(\alpha)z_1! \ldots z_r!}
      p_{0,i}^{\alpha} \; p_{1,i}^{z_1} \ldots p_{r,i}^{z_r}
    \end{align*}

    The log-likelihood is thus proportional to

    \begin{align*}
      \log\Gamma(\alpha+z_1+\ldots+z_r) &- \log\Gamma(\alpha) + \\
      \alpha\log(p_{0,i}) &+ z_1 \log(p_{1,i}) + \ldots +
      z_r \log(p_{r,i})
    \end{align*}

    The third term of (\ref{QEM}) is then (up to an additive
    constant)

    \begin{align}
      \ell = &-n \log\Gamma(\alpha) + \sum_{k=1}^n
      \log\Gamma(\alpha+z_{k,1}+\ldots+z_{k,r}) + \\
      &\sum_{i=1}^m\sum_{k=1}^n \phi_{k|n}(i) \Big(\alpha\log(p_{0,i}) +
      z_{k,1}\log(p_{1,i}) + \ldots + z_{k,r}\log(p_{r,i}) \Big).
\label{expl_nm}
    \end{align}

    The maximum is found by differentiation as shown below. We
    start by differentiating with respect to the parameters
    $p_{0,i}, \ldots, p_{r,i}$, which are bound by the constraint
    $p_{0,i} + \ldots + p_{r,i} = 1$.

    \begin{align}
      \frac{\partial \ell}{\partial p_{0,i}} = 
      \frac{\alpha}{p_{0,i}} \sum_{k=1}^n \phi_{k|n}(i)
      = \lambda
\label{p0}
    \end{align}

    \begin{align}
      \frac{\partial \ell}{\partial p_{l,i}} = 
      \frac{1}{p_{l,i}} \sum_{k=1}^n \phi_{k|n}(i) z_{k,l} = \lambda,
    \; l = 1, \ldots, r.
\label{p1}
    \end{align}

    The solution of equations (\ref{p0}) and (\ref{p1}) is given by

    \begin{align}
    p_{0,i} &= \frac{\alpha}{\alpha + \bar{z}_{1,i} +
      \ldots + \bar{z}_{r,i}} \label{sol_p0} \\
    p_{l,i} &= \frac{\bar{z}_{l,i}}{\alpha + \bar{z}_{1,i} +
      \ldots + \bar{z}_{r,i}}, \; \text{where} \\
    \bar{z}_{l,i} &= \frac{\sum_{k=1}^n\phi_{k|n}(i)z_{l,i}}
      {\sum_{k=1}^n\phi_{k|n}(i)}\; (l = 1, \ldots, r).
    \end{align}

    Equation (\ref{sol_p0}) is then used to obtain an equation in
    $\alpha$ by substitution.

    \begin{align}
      \frac{\partial \ell}{\partial \alpha}
      &= -n\psi(\alpha) + \sum_{k=1}^n 
      \psi(\alpha+z_{k,1}+\dots+z_{k,r})
      +\sum_{i=1}^m\sum_{k=1}^n \phi_{k|n}(i) \log(p_{0,i}) \nonumber \\
      &= n(\log(\alpha) - \psi(\alpha)) + \sum_{k=1}^n
      \psi(\alpha+z_{k,1}+\dots+z_{k,r}) \nonumber \\
      &- \sum_{i=1}^m \log(\alpha + \bar{z}_{1,i} +
      \ldots + \bar{z}_{r,i}) \sum_{k=1}^n\phi_{k|n}(i)
\label{lalpha}
    \end{align}

    The equation $\partial \ell / \partial \alpha \equiv f(\alpha) = 0$
    is solved by the Newton-Raphson method. For this we need to
    use the update formula
    $\alpha^{(t+1)} = \alpha^{(t)} - f(\alpha^{(t)})/f'(\alpha^{(t)})$,
    which depends on $f'(\alpha)$ which is computed as show below.
  
    \begin{align*}
    f'(\alpha) &= n\left(1/\alpha-\psi'(\alpha) \right) +
      \sum_{k=1}^n\psi'(\alpha+y_k+z_{k,1}+\ldots+ z_{k,r}) \\
      &- \sum_{i=1}^m \frac{1}{\alpha + \bar{z}_{1,i} +
      \ldots + \bar{z}_{r,i}} \sum_{k=1}^n\phi_{k|n}(i)
    \end{align*}

  \section{Negative binomial mixture model}
\label{mixture}

    In a negative binomial mixture model, every observation is drawn
    from a finite set of negative binomial distributions. Here we
    will only consider the case of two distributions. More specifically
    we will consider that the observations are drawn from a negative
    binomial with parameters $(\alpha, p)$ with probability $\theta$
    and from a negative binomial with parameters $(\alpha, q)$
    with probability $1-\theta$. The distribution is thus

    \begin{equation}
    P(Y = y) = \frac{\Gamma(\alpha+y)}{\Gamma(\alpha)y!}
    \left(\theta p^{\alpha}(1-p)^y + (1-\theta)q^{\alpha}(1-q)^y\right).
    \end{equation}

    Mixture distributions are commonly fitted by the EM algorithm.
    We suppose that an unobserved variable $Z$ takes value 1 with
    probability $\theta$ and value 0 with probability $1-\theta$.
    Obivously, $Z$ indicates which of the two distributions the
    observation is drawn from. The full likelihood is 

    \begin{equation*}
    P(y, z) = \frac{\Gamma(a+y)}{\Gamma(\alpha)y!}
      \left(\theta^z p^{\alpha}(1-p)^y + (1-\theta)^{1-z}
      q^{\alpha}(1-q)^y\right).
    \end{equation*}

    This immediately leads to the observation that

    \begin{equation}
    P(Z=1 | Y=y) = \frac{\theta p^{\alpha}(1-p)^y}
      {\theta p^{\alpha}(1-p)^y + (1-\theta) q^{\alpha}(1-q)^y}.
    \end{equation}

    The E-step of the algorithm is to write the expected log-likelihood
    of the distribution with respect to the conditional distribution
    of $Z$. If we write $\theta_k = P(Z_k=1 | Y=y_k)$ and drop the
    constant term, this quantity is

    \begin{align}
    \ell = -n \log \Gamma(\alpha) &+ \sum_{k=1}^n
      \log \Gamma(\alpha + y_k) + \nonumber
    \theta_k (\log(\theta) + \alpha \log(p) +
    y_k \log(1-p)) + \nonumber \\
    &\;\; (1-\theta_k) (\log(1-\theta) + \alpha \log(q) +
    y_k \log(1-q)).
\label{sub}
    \end{align}

    The M-step is to maximize (\ref{sub}), which is done by
    differentiation. Introducing
    $\bar{y}_1 = \sum_{k=1}^n \theta_k y_k / \sum_{k=1}^n \theta_k$
    and $\bar{y}_0 = \sum_{k=1}^n (1-\theta_k) y_k /
    \sum_{k=1}^n (1-\theta_k)$, it is easily verified that at
    the optimum

    \begin{align*}
    \theta &= \frac{1}{n}\sum_{k=1}^n\theta_k \\
    p &= \frac{\alpha} {\alpha+\bar{y}_1} \\
    q &= \frac{\alpha} {\alpha+\bar{y}_0}.
    \end{align*}

    By substituting those values in $\partial \ell / \partial \alpha$,
    we obtain an expression $f(\alpha)$ that depends on $\alpha$ only

    \begin{equation*}
       f(\alpha) = n \left( (\log(\alpha) - \psi(\alpha) \right) +
      \sum_{k=1}^n \psi(\alpha+y_k) + \theta_k \log(\alpha + \bar{y}_1)
      + (1-\theta_k) \log(\alpha + \bar{y}_0)
    \end{equation*}

    We need to find the solution of $f(\alpha) = 0$, which is done by
    the Newton-Raphson method. For this, we use the update formula
    $\alpha_{i+1} = \alpha_i - f(\alpha_i)/f'(\alpha_i)$, where

    \begin{equation*}
      f'(\alpha) = n \left( 1/\alpha - \psi'(\alpha) \right) +
      \sum_{k=1}^n \psi'(\alpha+y_k) +
      \frac{\theta_k}{\alpha+\bar{y}_1} +
      \frac{1-\theta_k}{\alpha + \bar{y}_0}.
    \end{equation*}

\noindent
    \textbf{Summary of the jahmm EM algorithm:} \par
    Assuming that the initial parameter values
    $\alpha_0, \theta_0, p_0, q_0$ are available, do the following:

    \begin{enumerate}

      \item For $k = 1, \ldots , n$ compute
        \begin{equation*}
        \theta_k = \frac{\theta_t p_t^{\alpha_t}(1-p_t)^{y_k}}
        {\theta_t p_t^{\alpha_t}(1-p_t)^{y_k} + (1-\theta_t)
         q_t^{\alpha_t}(1-q_t)^{y_k}}.
        \end{equation*}

      \item Compute
        \begin{align*}
        \bar{y}_1 &= \frac{\sum_{k=1}^n \theta_k y_k}
          {\sum_{k=1}^n \theta_k}, \\
        \bar{y}_0 &= \frac{\sum_{k=1}^n (1-\theta_k) y_k}
          {\sum_{k=1}^n (1-\theta_k)}. \\
        \end{align*}

      \item Update $\alpha$ by the Newton-Raphson scheme. Starting
        with $\tilde{\alpha}_{0} = \alpha_t$,
        update the value of $\tilde{\alpha}$
        with the formula
        $\tilde{\alpha}_{i+1} = \tilde{\alpha}_i -
        f(\tilde{\alpha}_i)/f'(\tilde{\alpha}_i)$, where
        \begin{align*}
          f(\tilde{\alpha}_i) &= n \left( (\log(\tilde{\alpha}_i) -
          \psi(\tilde{\alpha}_i) \right) + \\
          &\sum_{k=1}^n \psi(\tilde{\alpha}_i+y_k) +
          \theta_k \log(\tilde{\alpha}_i + \bar{y}_1)
          + (1-\theta_k) \log(\tilde{\alpha}_i + \bar{y}_0),
          \; \text{and} \\
          f'(\tilde{\alpha}_i) &= n \left( 1/\tilde{\alpha}_i -
          \psi'(\tilde{\alpha}_i) \right) +
          \sum_{k=1}^n \psi'(\tilde{\alpha}_i+y_k) +
          \frac{\theta_k}{\tilde{\alpha}_i+\bar{y}_1} +
          \frac{1-\theta_k}{\tilde{\alpha}_i + \bar{y}_0}.
        \end{align*}
        Stop iterations when
        $|\tilde{\alpha}_{i+1} - \tilde{\alpha}_i| < \varepsilon$ for
        a chosen $\varepsilon$, and set $\alpha_{t+1} =
        \tilde{\alpha}_{i+1}$.

      \item Update $\theta$, $p$ and $q$ by
        \begin{align*}
        \theta_{t+1} &= \frac{1}{n}\sum_{k=1}^n\theta_k \\
        p_{t+1} &= \frac{\alpha_{t+1}} {\alpha_{t+1}+\bar{y}_1} \\
        q_{t+1} &= \frac{\alpha_{t+1}} {\alpha_{t+1}+\bar{y}_0}.
        \end{align*}

      \item If the values of $\alpha$, $\theta$, $p$ and $q$ are
      stable stop the algorithm, otherwise start another cycle.
    \end{enumerate}

    \section{Negative multinomial mixture emissions}

    The parameters $\alpha$ and $\theta$ can be estimated from the
    reads counts of the negative control $(y_1, \ldots, y_n)$ by
    the EM algorithm as shown in section \ref{mixture}. We now
    turn to the Baum-Welch algorithm under the assumptions that
    the observations in each profile are drawn from a negative
    binomial mixture distribution of which the parameters $\alpha$
    and $\theta$ are the same.

    Dropping the constants terms (also including $\alpha$ which is
    now fixed), expression (\ref{expl_nm}) is replaced by

    \begin{align*}
    \ell = \sum_{i=1}^m\sum_{k=1}^n \phi_{k|n}(i)
      \log\Big(\theta p_{0,i}^{\alpha} \, p_{1,i}^{z_{k,1}} \ldots 
      p_{r,i}^{z_{k,r}} +
      (1-\theta) q_{0,i}^{\alpha} \, q_{1,i}^{z_{k,1}} \ldots
      q_{r,i}^{z_{k,r}} \Big).
    \end{align*}

    For simplicity, we introduce the terms $\theta_k(i)$ for
    $k = 1, \ldots, n$ and $i = 1, \ldots, m$ defined by
  
    \begin{align}
    \theta_k(i) = \frac{\theta p_{0,i}^{\alpha} \, p_{1,i}^{z_{k,1}}
    \ldots p_{r,i}^{z_{k,r}}}
    {\theta p_{0,i}^{\alpha} \, p_{1,i}^{z_{k,1}}
    \ldots p_{r,i}^{z_{k,r}} + (1-\theta) q_{0,i}^{\alpha} \,
    q_{1,i}^{z_{k,1}} \ldots q_{r,i}^{z_{k,r}}},
\label{thetaki}
    \end{align}

\noindent
    and the terms $\bar{z}_{l,i}^*$ for $l = 1, \ldots, r$ and
    $i = 1, \ldots, m$ defined by

    \begin{align*}
    \bar{z}_{l,i|1}^* &= \frac{\sum_{k=1}^n\phi_{k|n}(i)
      \theta_k(i)z_{l,i}}{\sum_{k=1}^n\phi_{k|n}(i)\theta_k(i)}, \\
    \bar{z}_{l,i|0}^* &= \frac{\sum_{k=1}^n\phi_{k|n}(i)
      (1-\theta_k(i))z_{l,i}}{\sum_{k=1}^n\phi_{k|n}(i)(1-\theta_k(i))}.
    \end{align*}

    Using a similar strategy as the EM, we can fix the $\theta_k(i)$
    and treat them as constants. The solution is subject to the
    constaints $p_{0,i}+\ldots+p_{r,i} = 1$,
    $q_{0,i}+\ldots+q_{r,i} = 1$, $p_{0,i}/p_{1,i} = C_1$ and
    $q_{0,i}/q_{1,i} = C_2$. Using Lagrange multipliers, we easily
    find that

    \begin{align*}
    p_{0,i} &= \frac{C_1}{C_1+1} \cdot \frac{\alpha+\bar{z}_{1,i|1}^*}
      {\alpha + \bar{z}_{1,i|1}^* +
      \ldots + \bar{z}_{r,i|1}^*}, \\
    p_{1,i} &= \frac{1}{C_1+1} \cdot \frac{\alpha+\bar{z}_{1,i|1}^*}
      {\alpha + \bar{z}_{1,i|1}^* +
      \ldots + \bar{z}_{r,i|1}^*}, \\
    p_{l,i} &= \frac{\bar{z}_{l,i|1}^*}{\alpha + \bar{z}_{1,i|1}^* +
      \ldots + \bar{z}_{r,i|1}^*}, \; (l = 2, \ldots, r).
    \end{align*}

\noindent
    and
    \begin{align*}
    q_{0,i} &= \frac{C_2}{C_2+1} \cdot \frac{\alpha+\bar{z}_{1,i|0}^*}
      {\alpha + \bar{z}_{1,i|0}^* +
      \ldots + \bar{z}_{r,i|0}^*}, \\
    q_{1,i} &= \frac{1}{C_2+1} \cdot \frac{\alpha+\bar{z}_{1,i|0}^*}
      {\alpha + \bar{z}_{1,i|0}^* +
      \ldots + \bar{z}_{r,i|0}^*}, \\
    q_{l,i} &= \frac{\bar{z}_{l,i|0}^*}{\alpha + \bar{z}_{1,i|0}^* +
      \ldots + \bar{z}_{r,i|0}^*}, \; (l = 2, \ldots, r).
    \end{align*}

    The new values of $\theta_k(i)$ are then recomputed by
    formula (\ref{thetaki}). Those EM-like cycles are repeated
    until convergence.

\end{appendices}

\end{document}